\documentclass[a4paper,12pt, parskip=full]{scrartcl}
\usepackage[T1]{fontenc}                  
\usepackage[utf8]{inputenc}               
\usepackage{textcomp}
\usepackage{lmodern}                      
\usepackage{setspace}                     
\usepackage{sectsty}                      
\usepackage{natbib}                       
\usepackage[protrusion=true,expansion=true]{microtype} 
\usepackage{booktabs}                     
\usepackage{graphicx}                     
\usepackage{amsmath}                      
\usepackage{latexsym}                     
\usepackage{bm,array}                     
\usepackage[labelfont=bf]{caption}        
\usepackage{authblk,etoolbox}             
\usepackage{algorithm}                    
\usepackage[noend]{algpseudocode}         
\usepackage{authblk}
\makeatletter
\patchcmd{\@maketitle}{center}{flushleft}{}{}
\patchcmd{\@maketitle}{center}{flushleft}{}{}
\def\maketitle{{%
  \renewenvironment{tabular}[2][]
    {\begin{flushleft}}
    {\end{flushleft}}
  \AB@maketitle}}
\def\BState{\State\hskip-\ALG@thistlm}
\makeatother

\DeclareUnicodeCharacter{00A0}{ }

\title{\textbf{Decline of war or end of positive check? }}
\subtitle{Analysis of change in war size distribution between 1816--2007}
\author{Stijn van Weezel}
\affil{Centre for International Conflict Analysis and Management, Nijmegen School of Management, Radboud University}
\date{ }

\begin{document}
\maketitle

\textbf{Abstract:}
\noindent
This study examines whether there has been a decline in the risk of death by battle during wars, testing the 'long peace' hypothesis. 
The analysis relies on the Expanded War Dataset \citep{gleditsch2004revised} covering inter- and inter-state wars between 1816--2007. 
Using untransformed data on war sizes, the estimates do not provide empirical evidence for a decline in war over time. 
However, normalising the data for global human population does illustrate a likely decline in war from 1947 onwards. 
The results indicate that despite strong population growth wars have not become more severe.

\noindent
{\Huge W}ether war has declined in recent history has been been the subject of much, and intense, debate recently. 
Research suggests that fewer wars are breaking out while more are terminating \citep{goldstein2011winning}, and that violence in general has been on the decline \citep{pinker2011better}. 
Indeed, since the end of the Second World War (1939--45) there has been an apparent decline in the risk of death by battle \citep{lacina2005monitoring,lacina2006declining}; a trend that does not seem to be abating \citep{pettersson2018organized,pettersson2019organized}. 
The relatively peaceful period in contemporary history --- from 1945 onwards --- has been dubbed the 'long peace' \citep{gaddis1986long}.
A term coined somewhat optimistically just four decades after the ending of one of the most cataclysmic events in recorded human history.

Some studies challenge the idea of the 'long peace', arguing there is no empirical evidence to support this claim based on the analysis of data on interstate wars \citep{clauset2018trends} and extremely large wars \citep{cirillo2016statistical} --- with statistical inference based on the fat-tailed distribution of wars sizes \citep{richardson1948variation,cederman2003modeling,gonzalez2016war}. 
Indeed, the 'long peace' can be criticised for being somewhat Eurocentric. 
Although violent armed conflict has been reduced greatly across the European continent, during the Cold War (1946--89) other regions in the world suffered greatly under the strain of war: Africa as a result of decolonisation \citep{arnold2017africa}; and most notably Asia, which was arguably the Cold War's main battlefield, with major powers pitted against each other, both directly and indirectly \citep{chamberlin2018cold} --- see \citet{wimmer2009location} for an exposition on war location.\footnote{From a historical perspective war has been on the wane in Central- and South-America \citep{bates2007lost}. 
But while the proportion of war (measured relative to the total number of wars globally) has increased in Asia in recent years \citep{aas2011all}, parts of the continent have experienced a substantial decline in warfare \citep{svensson2011bombs} (e.g. Bangladesh, Indonesia, Malaysia, Viet Nam). For a broader discussion on modern war see \citet{wimmer2006empire}.}

Studies refuting the 'long peace' or decline-of-war hypothesis clash with empirical regularities in the data identified by other work which does illustrate a decreasing trend in war as a function of time \citep{martelloni2018pattern}. 
Recent contributions to this debate have focused predominantly on identifying changepoints in the war size distribution; assessing points in time after which the distribution changed indicating a decline in war (measured by the number of battle-deats). 
Using this approach 1950 and 1965 have been identified as potential changepoints for interstate wars \citep{cunen2018statistical}, while other studies, using a broader sample of wars, have identified potential changepoints around 1950 \citep{fagan2019changepoint,spagat2020decline}, with the 1830s, 1910, and 1994 as other contenders \citep{fagan2019changepoint}.

The study presented in this paper is similar in spirit to the changepoint-oriented studies. 
But whereas the existing work employs complex statistical methods, and focuses predominantly on war sizes, this study takes a simpler approach and focuses on the risk of death in battle.  
To test whether there has been a decline in war, and when this potential decline started, the analysis relies on inference of the binomial proportion of war sizes. 
Specifically, the frequency of wars of a particular size can be measures as a proportion of the total number of wars, both before and after a designated changepoint. 
This proportion can be straightforwardly modeled as a set of Bernoulli trials using the binomial model. 
Bayesian inference is used to estimate the distribution of the parameters. 
This approach provides the advantage of producing estimates which allow a probabilistic interpretation. 
This in contrast with quantities of interest obtained through orthodox hypothesis tests with accompanying $p$-values (e.g. \citet{spagat2020decline}).

Using a comprehensive dataset covering wars between 1816--2007 \citep{gleditsch2004revised}, the analysis provides some empirical evidence for a decline in war; specifically from 1947 onwards.  
The estimated probability that the risk of being killed in battle is lower between 1947--2007 compared to 1816---1946 is at least 0.66; a threshold at which we can state with some certainty that a decline has occurred \citep{mastrandrea2010guidance}. 
However, it is important to note that this decline is only apparent when war sizes are normalised for the total global population (to factor in the average risk of someone being killed in battle).
Although the raw untransformed data exhibits similar patterns, the estimated probabilities are much lower. 
As such, these provide no convincing evidence in favour of the decline-of-war hypothesis, but neither do they suggest that things have gotten worse. 

There are some other important caveats concerning the analysis of the results and the interpretation. 
There is the data quality, particularly the measurement of the number of battle-deaths. 
It is very plausible that the reported number of deaths are measured with substantial error; most importantly they potentially measure different concepts for different wars, mixing direct and indirect deaths \citep{lacina2005monitoring}. 
Although other studies acknowledge the measurement error they tend to be silent with regard to the concept that enters the analysis; using the data as-is. 
The practical implication of this is that it is actually hard to make like-for-like comparisons without the analysis relying on some heroic assumptions. 
For this particular study that the estimated war sizes are not off by more than half an order of magnitude. 
In addition, although there has been a substantial increase in global population over time \citep{roser2013world}, this has not corresponded, so it seems, to an increase in war severity. 
Some have pointed to a possible explanation in improvements in military medicine \citep{fazal2014dead}, although this account seems to neglect other, more long-term, technological advances linked to mobilization levels and army size \citep{onorato2014technology} which are strongly linked to fatality numbers \citep{oka2017population}.

\section*{Data}
\begin{figure}[!ht]\centering
  \includegraphics[scale=.5]{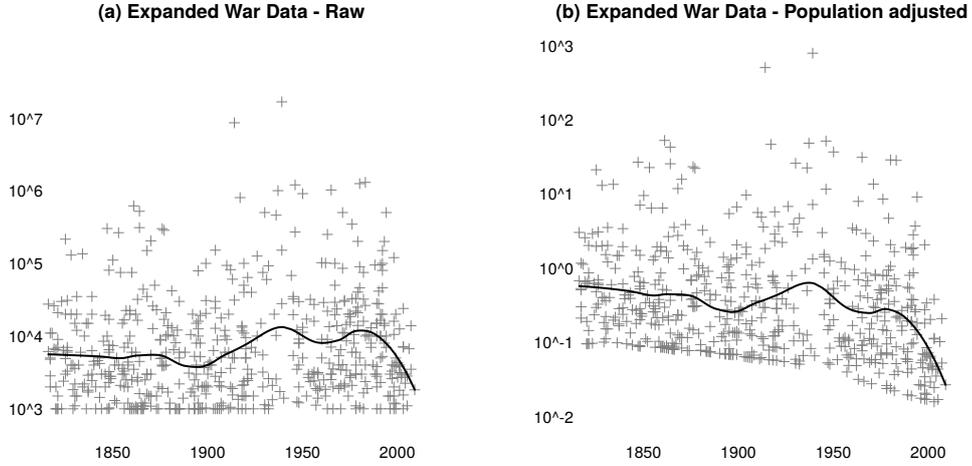}
  \caption{Expanded War Dataset: Distribution of war sizes over time (1816--2007) smoothed with loess smoother with span of one third (solid line).
  \newline \textbf{(a)} original data; \textbf{(b)} population-normalised data.
  \newline \textit{Data source:} \citet{gleditsch2004revised}.}
  \label{fig:data}
\end{figure}

Information on wars is taken from the Expanded War Dataset (EWD) compiled by \citet{gleditsch2004revised}.
This dataset covers both intra- and inter-state wars between 1816--2007 ($N=570$).
The analysis includes both war types as there is no theoretical justification to analyse them in separation \citep{cunningham2013combining} or focus exclusively on interstate wars as others have done. 
Excluding civil wars could potentially lead to erroneous conclusions since this type of war has been increasing over time \citep{miranda2016evolution}.
Although the EWD is a comprehensive dataset, it likely suffers from some measurement error.
Most notable caveat of the dataset is that it only includes war with a minimal size of $10^3$ battle-deaths (64 wars actually have this specific size).
As such, smaller wars are excluded from the analysis, which could bias the results against the decline-of-war hypothesis. 
In addition, there could be a bias arising from reporting error, both in terms of the number of battle-deaths as well as the number of wars itself (indeed, the EWD is preferred over the commonly used Correlates of War data as it addresses a number of biases arising from coding decisions \citep{gleditsch2004revised}).
The extent of this error is unknown and deserves more attention in future research. 
Finally, there is the issue that the number of battle-deaths might represent different concepts for different wars as discussed by \citep{lacina2006declining}.

War size $w_{it}$ is measured as the total number of battle-related deaths for the entire duration of war $i$. 
To approximate the risk of dying during battle, the war size is normalised using total world population taken at the year of onset $t$ with $1816 \leq t \leq 2007$.  
Hence normalised war size $\widetilde{w}_{it}$ equals $\frac{w_{it}}{pop_t}$. 
To create an annual time-series of world population a number of different data sources are used \citep{klein2010long,vanzanden2014global,un2017world}.
Missing values, most common before 1950, are linearly interpolated. 
Figure~\ref{fig:data} displays the data; table~\ref{table:summary-statistics} provides summary statistics.

\begin{table}[!ht]\centering
      \caption{Summary statistics ($N=570$)}
  \label{table:summary-statistics}
    \begin{tabular}{lccccc}
    \\[-1.8ex]\hline \hline \\[-1.8ex]
    ~                   & Mean                               & Median             & Range                 & Skewness & Tail$^a$ \\ \hline \\[1.8ex]
    War size            & 0.8 $\cdot 10^5$ (8 $\cdot 10^5$)  & 0.04 $\cdot 10^5$  & $[10^3; 2 \cdot10^7]$ & 18.2     & 0.086 \\
    War size normalised & 0.4 $\cdot 10^1$ (4 $\cdot 10^1$)  & 0.03 $\cdot 10^1$  & $[0.02; 781]$         & 17.3     & 0.074 \\
    \\[-1.8ex]\hline  \\[-1.8ex]
    \multicolumn{6}{p{25em}}{\emph{Note:} Standard deviation (between parentheses)
    \newline $a$: Proportion of observations that is larger than the mean.}
    \end{tabular}
\end{table}

\section*{Methods}
\begin{figure}[!ht]\centering
  \includegraphics[scale=.5]{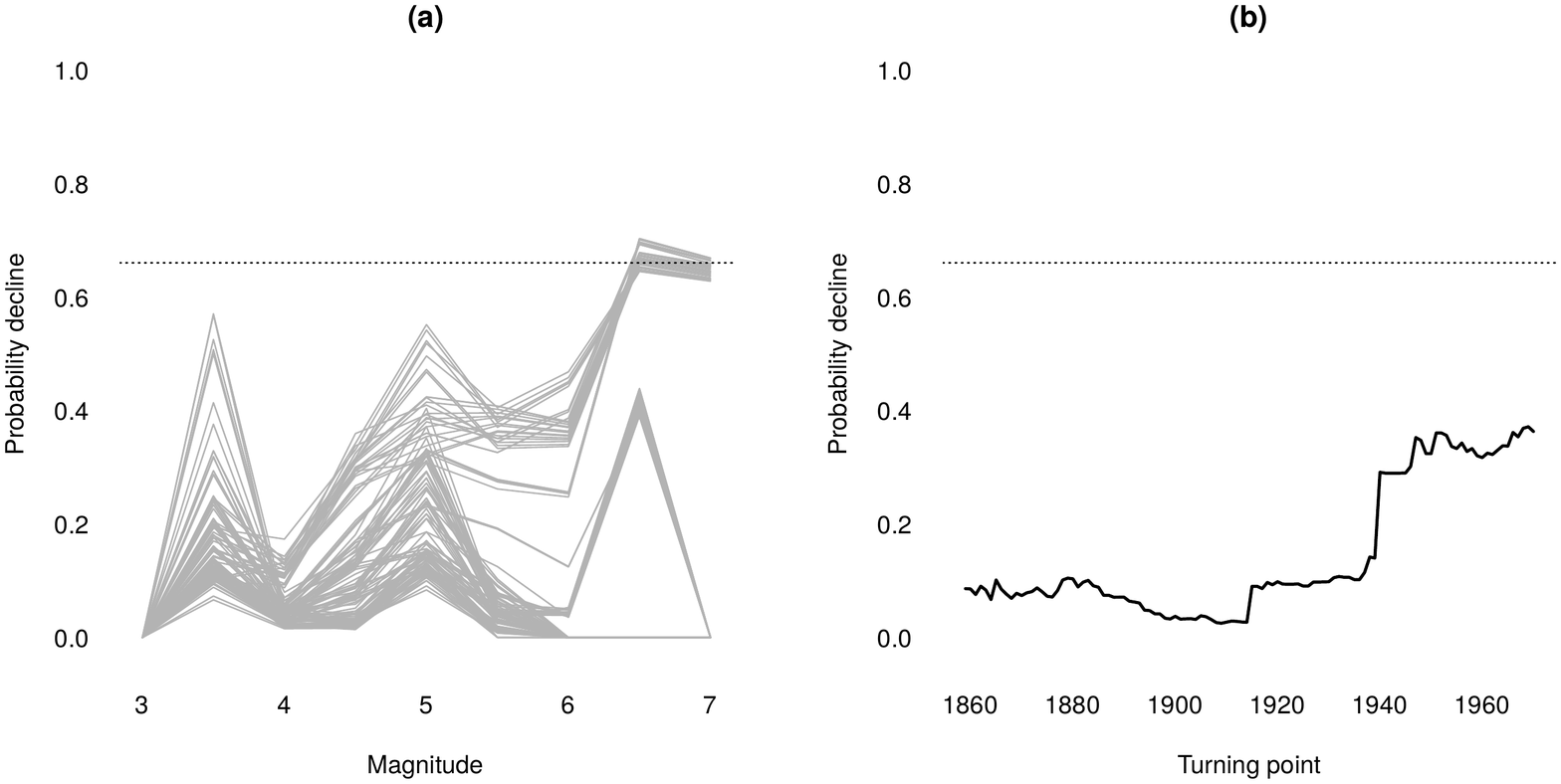}
  \caption{Conflict onset frequency.
  \newline \emph{Note:} \textbf{(a)} Number of conflict onsets per year. The dotted line indicates the sample average ($\sim$3), the solid line denotes a 5-year moving average; \textbf{(b)} minimal number of years between conflicts. Solid circles indicate the observed proportions, the line indicates the expected proportions based on the average of $10^3$ random Poisson distributions: $Pois(\lambda)$ with $\lambda$ set equal to the sample average (0.34).  
  \textit{Data source:} \citet{gleditsch2004revised}.}
  \label{fig:poisson}
\end{figure}

The quantity of interest is the frequency of wars with minimal magnitude $m$; both before and after potential changepoint $\hat{t}$.
This frequency can be measured as a proportion relative to the total number of wars for each period and the binomial model can be used to model wars with size $\geq m$ as a set of Bernoulli trials $X_i, ($i=1,...,n$)$.
The underlying assumptions is that the wars are independent of each other, in other words that the war-generating process is memoryless. 
As such, no time trend is assumed, instead a stationary process governs the frequency and distribution of war sizes \citep{clauset2020frequency}. 
This does not seem to be a too heroic assumption given the relative constant rate of war outbreaks over time, which seems stationary around a mean (figure~\ref{fig:poisson}\emph{a}). 
In addition the time between wars follows a random Poisson process \citep{clauset2018trends} as illustrated by figure~\ref{fig:poisson}\emph{b}.\footnote{This assumption also implies there is such a thing as a common global war-generating process. Whether this assumption is tenable is up for debate, and when violated could risk the analysis being subject to an ecological fallacy. The 'long peace' hypothesis does rely on this assumption as the decline in war is linked to a global pacifying process where diffusion of norms and values, for instance through trade and increased connectivity between communities, make societies less bellicose.} 

Let $Y$ be the total observed number of wars with size $\geq m$ and $n$ the total number of wars.\footnote{This section draws from \citet{bolstad2016introduction}.}
The probability of war $y$ being of size $\geq m$ can be denoted by $\theta$ (ie. the frequency). 
$Y$ follows a binomial distribution: $Y \sim B(n, \theta)$ which entails that the conditional probability of $y$ based on $\theta$ is given by:  

\begin{align}
   p(y | \theta) = \binom{n}{y} \theta^y (1-\theta)^{n-y} \; \text{for} \; y = 0,1,...,n
\end{align} 

Within this frameworks inference of $\theta$ is based on the observed proportion $p=y/n$ which serves as a point estimate. 
The likelihood function can be written as: 

\begin{align}
  \mathcal{L}(\theta) \propto p(y | \theta)  
\end{align}

where $\mathcal{L}$ can be estimated using maximum likelihood estimation (MLE) to obtain the most likely value for $\theta$ given the data. 

Testing whether there has been a decline in war, of interest is whether the value of $\theta$ has changed over time, specifically between two periods. 
To this end orthodox tests of statistical significance could be employed, but these come with two important caveats. 
First, they rely on asymptotic properties which can be problematic in small samples. 
Second, they only offer a binary answer --- to reject or accept the null hypothesis --- rather than information about the probability with which a certain development has occurred. 
Therefore this analysis used Bayesian inference to circumvent these issues. 
In this context and additional advantage of the Bayesian approach is that prior information, or existing beliefs, about the frequency of wars can easily be incorporated in the analysis using Bayes' Theorem. 
Here $\theta$ is conditional on outcome $y$ (and $n$ which is omitted for ease of notation) with the posterior distribution for $\theta$ given by: 

\begin{align}
  p(\theta | y) &= \frac{p(\theta) p(y | \theta)}{p(y)}, 
  \text{with}\;p(y) = \int p(y | \theta) p(\theta) d\theta
\end{align}

$p(y)$ is a scaling factor denoting the marginal density of the data such that the probabilities sum to one. 
Omitting this scaling factor Bayes' Theorem can be rewritten in a simplified form: 

\begin{align}
 p(\theta | y) \propto p(y | \theta)p(\theta) 
\end{align}

In plain English this formulation states that the posterior distribution is proportional to the likelihood times the prior distribution. 
As already established the likelihood of $\theta$ can be described by the Bernoulli distribution which only leaves the prior to be specified.  
Since $\theta$ is defined on the unit interval ($0 < \theta < 1$) the Beta distribution --- $Beta(a,b)$ --- would be a natural choice as it is bounded by the same interval. 
Indeed, the Beta distribution is a conjugate prior for the Bernoulli distribution, which entails that it will produce a Beta-distributed posterior; ie. defined on the unit interval.  

The Beta distribution also provides flexibility in specifying the its shape parameters ($a, b$), which is useful with regard to the prior distribution. 
Specifying a prior one could opt for using a diffuse prior assigning equal probability to different war frequencies. 
In that case the estimate for $\theta$ would be similar to what one would obtain using MLE.
But it would also ignore the available information on war frequencies, specifically the fact that the distribution of war sizes is fat-tailed \citep{richardson1948variation,cederman2003modeling,gonzalez2016war}.  
Hence, based on prior beliefs we would expect that the frequency of huge wars is smaller than the frequency of large wars which is smaller than the frequency of small wars. 
Therefore, with the prior distribution $p(\theta)$ specified as $\theta \sim Beta(a, b)$, the shape parameters are defined such that they capture existing beliefs. 
In this particular case $a$ is the number of wars of size $\geq m$ before the changepoint and $b$ the number of wars with smaller magnitudes (meaning the the total number of wars before the changepoint is given by $a + b$).
In practical terms this means that the posterior distribution of $\theta$ can be inferred by sampling from the Beta distribution with shape parameters set to $y+a$ and $n-y+b$ as illustrated by equation~\ref{eq:beta} below:

\begin{subequations}
\label{eq:beta}
\begin{align}
   p(\theta) &\propto \theta^{a-1} (1-\theta)^{b-1} \\
   \mathcal{L}(\theta) &= \theta^y (1-\theta)^{n-y} \\
   p(\theta | y) &\propto \mathcal{L}(\theta) p(\theta) \\ 
   &\propto  \theta^y (1-\theta)^{n-y}  \cdot \theta^{a-1} (1-\theta)^{b-1} \\ 
  &\propto \theta^{y+a-1} (1-\theta)^{n-y+b-1} \\ 
  & = Beta(y+a,\; n-y+b)  
 \end{align} 

\end{subequations}

The posterior is estimated using $10^4$ draws from the Beta distribution. 
The shape parameters are defined partitioning the data across time and war size. 
Using the time-series variation of the data, each year $t$ is considered a potential changepoint $\hat{t}$, with $1859 \leq \hat{t} \leq 1970$.
Admittedly this interval might seem somewhat arbitrary, but the boundaries are actually theoretically informed.
Research has illustrated an upward shift in military size and mobilization levels --- important determinants of war fatalities \citep{oka2017population} --- from 1859 onwards, followed by a decline from 1970 \citep{onorato2014technology}. 
These trends coincide with the advance of railways, facilitating higher mobilisation rates, and improvements in guided-missile technology, which are correlated with reductions in army size \citep{onorato2014technology}. 

Splitting the data at year $\hat{t}$ information from years $t \leq \hat{t}$ is used to specify parameters of the prior distribution --- $p(\theta): \theta \sim Beta(a, b)$.
The values for $a, b$ are determined by a second data partition based on war size $m$. 
Where $a$ is the number of wars with size $\geq m$ and $b$ the number of wars with size $< m$ (see the example of pseudo-code~\ref{alg:pseudo-code}).

Naturally the reported number of people killed by a particular war comes with a degree of uncertainty --- in addition to other biases previously discussed; therefore the war sizes are log-transformed to the base 10 \citep{richardson1948variation} and the data is split using a relatively small number of cut-off points. 
The data is unlikely to be fully accurate given the issues measuring war-related mortality \citep{osterud2008towards,spagat2009estimating,hacker2011census,jewell2018accounting}.

To gauge whether a decline in war has occurred a relatively simple measure is used: The difference between the prior and posterior distribution. 
If no change occurred before and after $\hat{t}$ then the prior distribution should be a good description of the war size distribution for the period following the changepoint. 
As such, the 'new' data would not nudge us towards adjusting the prior and hence the difference between prior and posterior should be negligible, close to zero on average. 
If there is indeed a decline in war --- here measured as the change in proportion of wars of a particular size --- the expectation is that the difference between posterior and prior is negative. 
The proportion of relatively larger wars should be higher for the prior then it is for the posterior. 

To translate this into a workable statistic the fraction of negative values relative to the total distribution is calculated. 
Recall that for the prior and posterior there are $10^4$ estimated values. 
These are subtracted from each other and the quantity of interest is the fraction of negative values. 
This measure has a probabilistic interpretation as it can be thought of as a probability of decline, where higher values correspond to a decline in wars with size $\geq m$ (see figure~\ref{fig:example} for an example).

\section*{Results}
\begin{figure}[!ht]\centering
  \includegraphics[scale=.5]{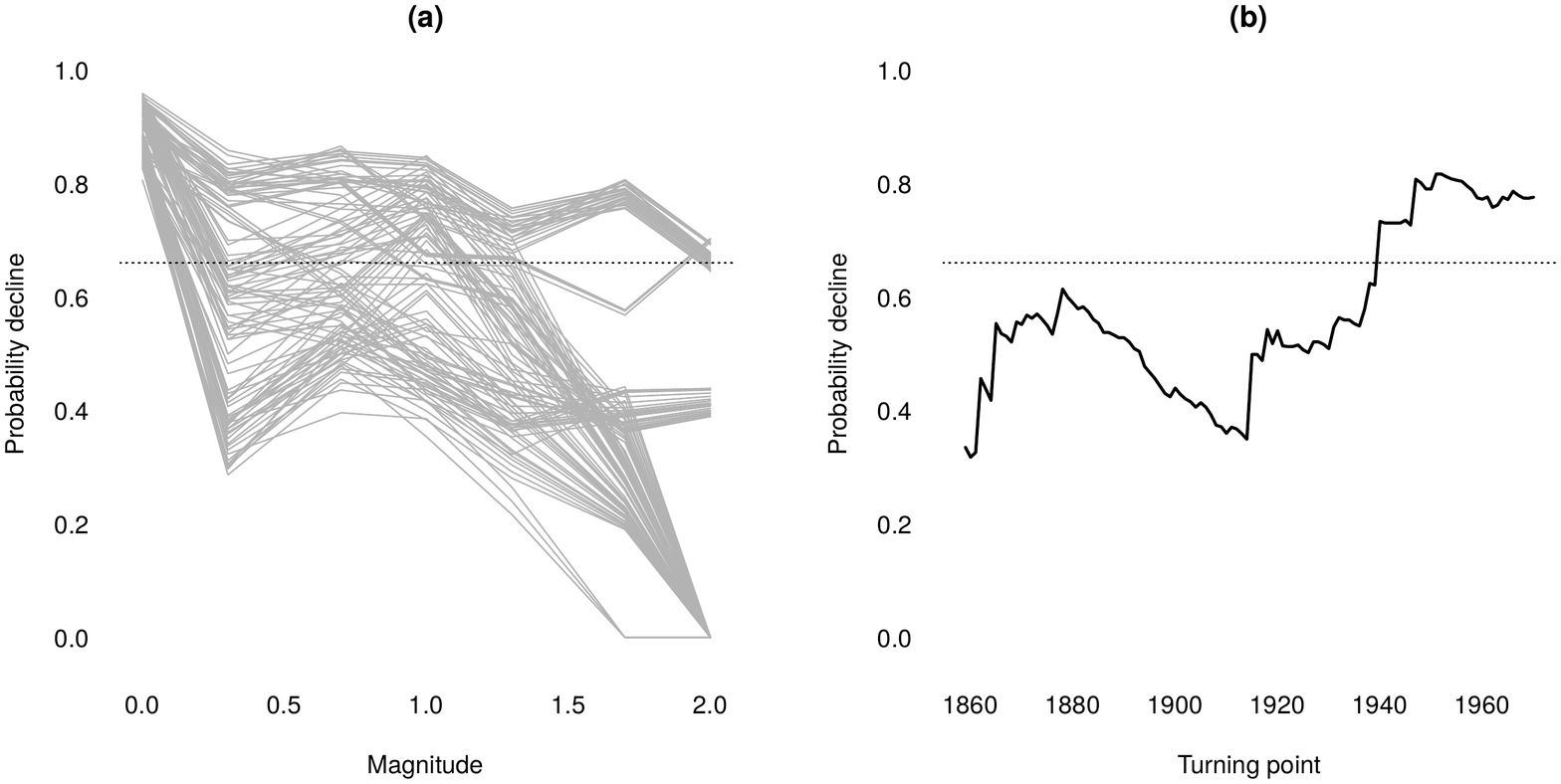}
  \caption{Decline in war severity --- estimated probability (raw data).   
  \newline \emph{Note:} \textbf{(a)} Probability of decline of war of size $\geq m$ for all potential turning points $\hat{t}$ with $1856 \leq \hat{t} \leq 1989$ denoted by light-grey lines; $m$ on interval $(3, 3.5,..., 7)$; \textbf{(b)} probability of decline average across all war sizes. 
  Solid horizontal line denotes probability of 0.5; dashed line probability of 0.66.
  \textit{Data source:} \citet{gleditsch2004revised}.}
  \label{fig:estimates-r}
\end{figure}

Figure~\ref{fig:estimates-r}\emph{a} summarises the results using the raw untransformed data, displaying the probability of decline for all possible changepoints across different magnitudes. 
Naturally the estimate is zero for war size $m \geq 3$ as this is the lower bound of the data; wars are included in the data if they attain at least $10^3$ battle-related deaths.
In general there is not much empirical evidence in favour of a decline-in-war type of hypothesis. 
Only at the left end of the interval do the results show that for a subset of years the probability of decline is relatively high, but only for war sizes $\sim m \geq 6$ (wars with a severity above $10^6$ are extremely rare).
Except for this particular range, the estimated probabilities rarely exceed the threshold of 0.66, indicating that a decline in war is as likely as no decline (quantifying uncertainty the guidelines of \citet{mastrandrea2010guidance} are followed). 
The estimates presented in panel (a) potentially obscure relevant temporal patterns. 
Therefore the estimated probabilities are averaged at annual level, as shown in figure~\ref{fig:estimates-r}\emph{b}. 
Note that an underlying assumption of this approach is that a decline in war is noticeable across the whole rage of magnitudes. 
This assumption does not hold entirely as the proportion of wars with size $m \geq 3$ will always be 1 due to the manner in which the data is collected. 
In addition it also entails that shifts in the war-size distribution are unaccounted for. 
The results show that across a range of potential changepoints a decline in war seems unlikely. 
Only following the Second World War does the probability indicating a change start to creep upward, but it remains low in absolute value peaking at 0.37. 

\begin{figure}[!ht]\centering
  \includegraphics[scale=.5]{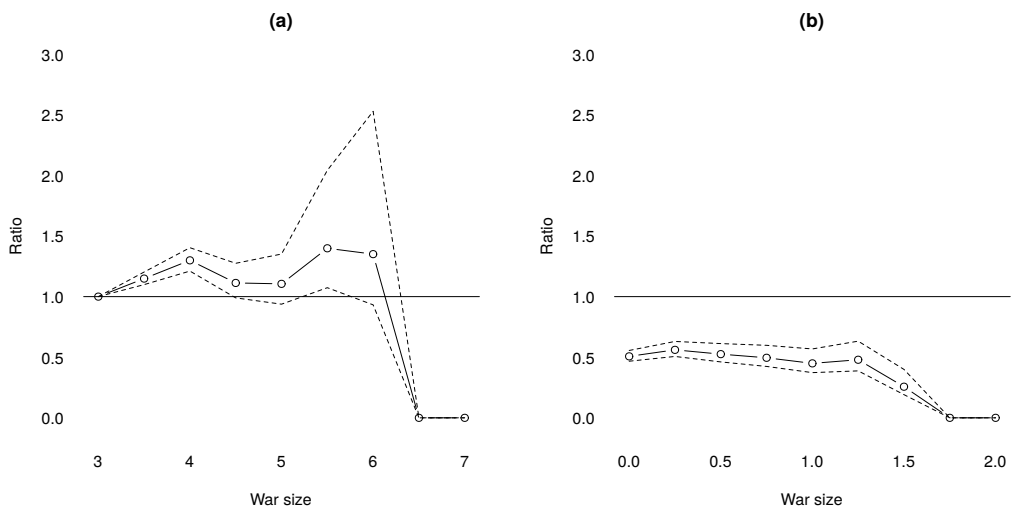}
  \caption{Decline in war severity --- estimated probability (normalised data).   
  \newline \emph{Note:} \textbf{(a)} Probability of decline of war of size $\geq m$ for all potential turning points $\hat{t}$ with $1856 \leq \hat{t} \leq 1989$ denoted by light-grey lines; $m$ on interval $(1,2,5,10,20,50,100)$; \textbf{(b)} probability of decline average across all war sizes. Solid horizontal line denotes probability of 0.5; dashed line probability of 0.66.
  \textit{Data source:} \citet{gleditsch2004revised}.}
  \label{fig:estimates-a}
\end{figure}

A caveat of the raw battle-deaths data is that it does not account for changes in the risk of people dying during a war \citep{lacina2006declining}. 
Therefore the data is normalised for global population; the results are shown in figure~\ref{fig:estimates-a}.
The results display a similar pattern to those reported in figure~\ref{fig:estimates-r}, but there is a noticeable shift upwards, along the $y$-axis, indicating higher probabilities of a change, favouring the decline-of-war hypothesis (Fig.~\ref{fig:estimates-a}\emph{a}). 
Surprisingly though this is a trend across different potential changepoints. 
For a broad range of war sizes the estimated probability of decline surpasses the 0.66 threshold. 
Averaging the probabilities, to get a single statistic for each year, shows an increase in the probability of decline following the First World War (1914--18) and around the Second World War (1939--45) (Note that the wars are indexed by their year of onset).
Indeed 1940 is the first in which the average estimated probability exceeds 0.66 after which it doesn't drop below this threshold.
Of note is that the estimated probability is also relatively high ($\geq 0.5$) during the period of European colonisation of Africa through the 1880-90s. 

\begin{figure}[!ht]\centering
  \includegraphics[scale=.5]{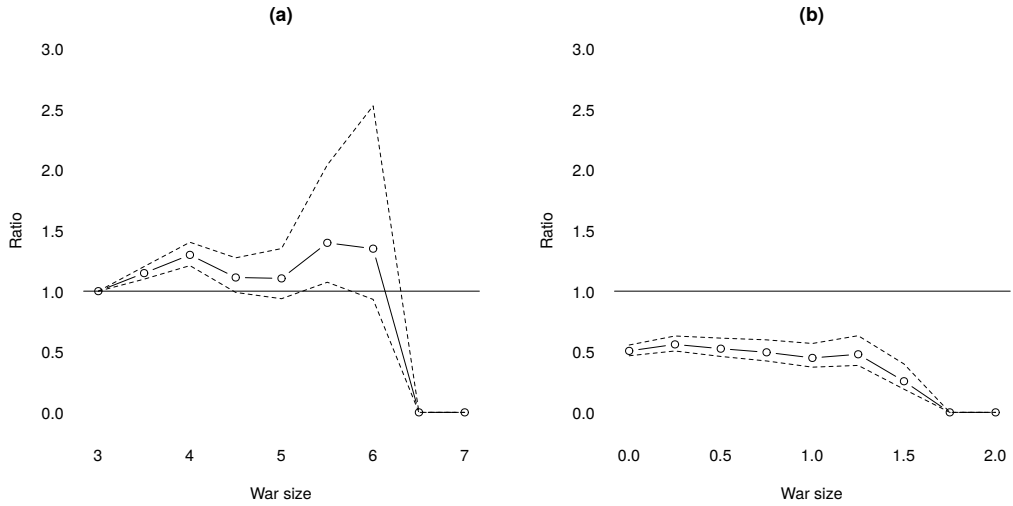}
  \caption{Predicted proportions: observed proportion of wars of size $\geq m$ relative to the predicted numbers using 1947 as turning point.
  \newline \emph{Note:} \textbf{(a)}, raw data; \textbf{(b)}, normalised data; dashed lines denote 66\% uncertainty interval.     
  \textit{Data source:} \citet{gleditsch2004revised}.}
  \label{fig:prediction}
\end{figure}

The estimates point to 1947 as the most likely changepoint; using the maximum of the estimated probabilities. 
Therefore an additional test is carried out partitioning the data at 1946 and focusing on the generated predictions of the prior distribution and how well they predict the proportion of wars between 1947--2007 (see pseudo-code~\ref{alg:pseudo-code2}). 
Similar to the previous test the observed proportions of the prior period (1816--1946) are used to generate predictions for the posterior period (1947--2007) using the Beta distribution. 
The average of $10^4$ draws from the prior distribution serves as point prediction, also accounting for the uncertainty associated with the estimate using an uncertainty interval.  
Figure~\ref{fig:prediction} plots the results for the raw (a) and normalised (b) data.  
The panels display the observed proportions relative to the predicted proportion, along with a 66\% uncertainty interval. 
Between 1946--2007 there is an apparent increase in wars with magnitudes ranging from $10^3$ to about $10^4.5$.
There is large uncertainty associated with the estimates for wars in the range of $10^5/10^6$ battle-deaths.
This is most probably the result of a small number of data points. 
The ratio drops to zero after $10^6$ as no wars of this magnitude have occurred since 1946, yet. 

Based on the raw battle-deaths data we would be inclined to draw the conclusion that there indeed has not been a noticeable decline in war severity across time. 
The data normalised for population provides conflicting evidence however. 
Importantly, taking into account the developments in population growth over the past two centuries --- which have not been hindered by the positive checks of war --- the historical pattern of war (ie. wars between 1816--1946) tends to over-predict war severity for the period hence. 
Even after accounting for the uncertainty in the generated predictions the result show that the observed proportions of war severity are around only half what would be expected, for most magnitudes, as illustrated by figure~\ref{fig:estimates-r}\emph{b}. 
There is a noticeable exception to this pattern, namely the peak at around 32 deaths per $10^5$ people. 
Here the observed proportions are more in line with expectations based on historical data. 

\section*{Conclusions}
The empirical analysis provides little empirical evidence in favour of the 'long peace' hypothesis when exclusively focusing on unadjusted battle-deaths data, consistent with existing work. 
The estimations show that the probability of a decline in war has increased over time, but at no point does it exceed any threshold indicating that a decline in war is more likely than no change in the assumed global war-generating process. 
The conclusions changes however when considering population-adjusted data to account for the average risk of being killed in battle during a war. 
Using this approach the results echo earlier studies on changepoint in war, indicating that the period after the Second World War does indeed seem to be relatively less belligerent in terms of adjusted fatality numbers. 
These results also illustrate that despite substantial population growth wars have not become more severe over time.

\let\oldbibliography\thebibliography
\renewcommand{\thebibliography}[1]{\oldbibliography{#1}
\setlength{\itemsep}{1pt}} 
\bibliographystyle{chicago}
\bibliography{references.bib}

\setcounter{figure}{0}
\renewcommand{\thefigure}{A\arabic{figure}}
\setcounter{table}{0}
\renewcommand{\thetable}{A\arabic{table}}
\section*{Appendix}

\begin{algorithm}
  \caption{Pseudo-code for estimating change in war severity} \label{alg:pseudo-code}
  \begin{algorithmic}[1]
    \State Set $N=10^4$
    \State Consider turning point $\hat{t}$ in $\{1859, 1970\}$     
    \State Define $\{k, n \}$ where $k$ and $n$ are number of wars before and after time $\hat{t}$, respectively    
    \State Consider magnitude $m$ in $range(m)$  
    \State Define $\{a, y\}$ where $a$ and $y$ are number of wars before and after time $\hat{t}$ with minimal size $m$, respectively       
    \State Sample $p(\theta) = Beta(N, a, k-a)$ 
    \State Estimate $p(\theta | y) = Beta(N, y+a, n-y + b)$ with $b = k-a$
    \State Calculate $D = p(\theta | y) - p(\theta)$
    \State Calculate $Pr(decline) = length(D[D < 0])/length(D)$ where $D[D < 0]$ are all the elements of $D$ with a value $<$0
    \State Repeat step 4--9 for each individual $m$ in $range(m)$  
    \State Repeat step 2--10 for each individual $\hat{t}$ in $\{1859: 1970 \}$ 
  \end{algorithmic}
\end{algorithm}

\begin{algorithm}
  \caption{Pseudo-code for predicting proportion of wars with minimal size $m$} \label{alg:pseudo-code2}
  \begin{algorithmic}[1]
  \State Set $N=10^4$
  \State Consider magnitude $m$ in $range(m)$
  \State Define $\{a, a+b\}$ where $a$ is total number of wars between 1816--1946 with minimal size $m$ and $a+b$ is total number of wars between 1816--1946
  \State Define $\{y, n\}$ where $y$ is total number of wars between 1947--2007 with minimal size $m$ and $n$ is total number of wars between 1947--2007
  \State Generate $\hat{p} = Beta(N, a, b)$
  \State Calculate observed proportion $p=y/n$  
  \State Calculate $\frac{p}{\hat{p}}$
  \State Repeat step 2--7 for each individual $m$ in $range(m)$
  \end{algorithmic}
\end{algorithm}

\begin{figure}[!ht]\centering
  \includegraphics[scale=.5]{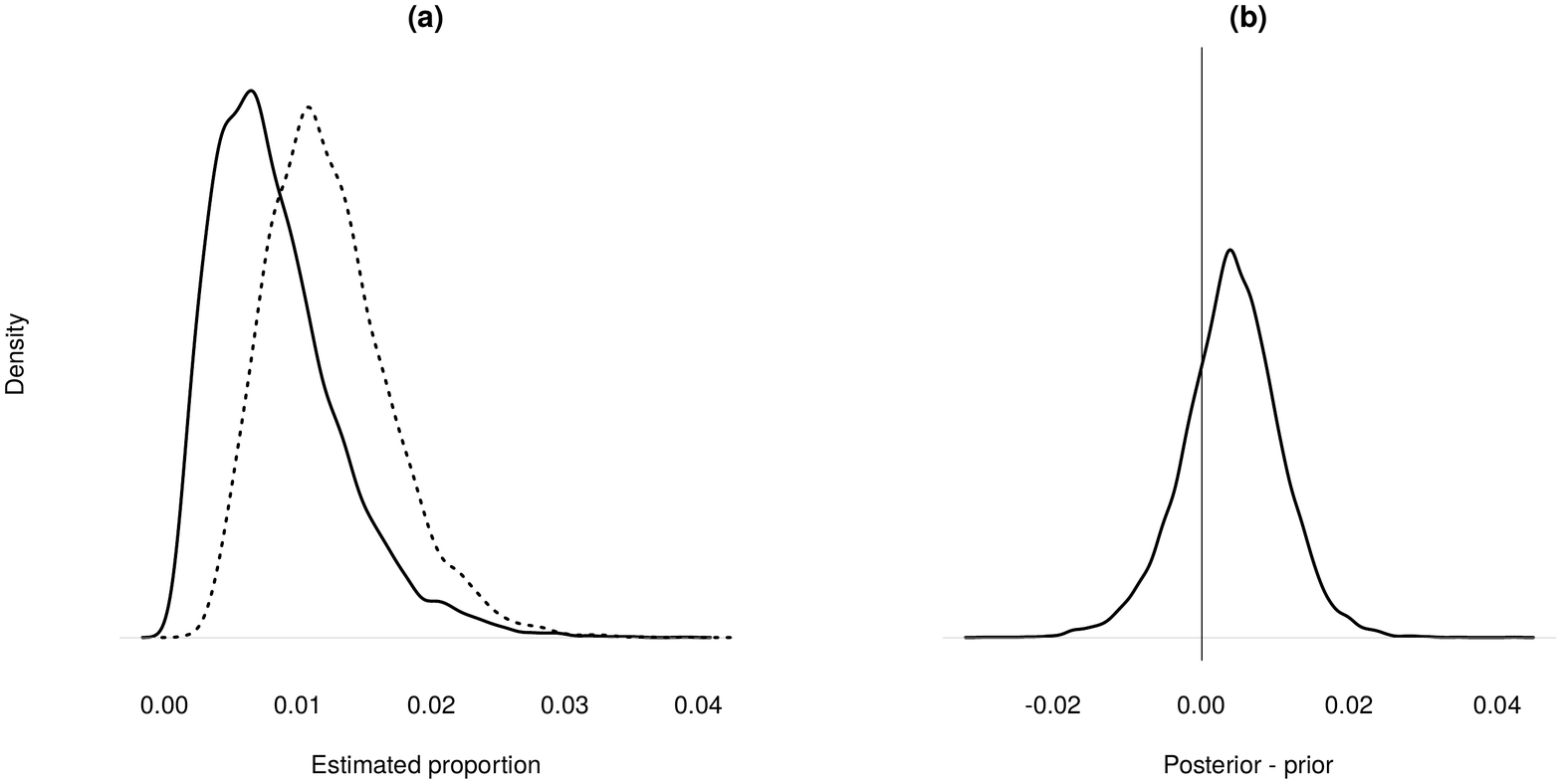}
  \caption{Example of applied method.
  \newline \textbf{(a)} displays the estimated posterior (dashed line) and prior (solid line) distribution for wars with size $\geq 10^6$ for the data partitioned at $t = 1950$. \textbf{(b)} illustrates there is seemingly little difference between the prior and posterior. Indeed, the probability of decline is only 0.25 --- indicating that an increase in wars with size $\geq 10^6$ is actually more likely.}
  \label{fig:example}
\end{figure}

\end{document}